\pdfoutput=1
\documentclass{JINST}
\let\ifpdf\undefined
\usepackage{textcomp} 
\DeclareGraphicsExtensions{.png,.pdf,.jpg} 
\usepackage{ifpdf}

\title{High speed readout electronics development for frequency-multiplexed kinetic inductance detector design optimization}

\author{O. Bourrion$^a$\thanks{Corresponding author},
C.~Vescovi$^a$,
A.~Catalano$^a$,
M.~Calvo$^b$,
A.~D'Addabbo$^b$,
J.~Goupy$^b$,
N.~Boudou$^b$,
J.F.~Macias-Perez$^a$ and A.~Monfardini$^b$.\\

\llap{$^a$}Laboratoire de Physique Subatomique et de Cosmologie,\\ 
Universit\'e Joseph Fourier Grenoble 1,\\
CNRS/IN2P3, Institut Polytechnique de Grenoble,\\
53, rue des Martyrs, Grenoble, France\\

\llap{$^b$}Institut N\'eel, CNRS/UJF, \\
25, rue des Martyrs, Grenoble, France \\}

\abstract{Microwave Kinetic Inductance Detectors (MKID) are a promising solution for space-borne mm-wave astronomy. To optimize their design and make them insensitive to the ballistic phonons created by cosmic-ray interactions in the substrate, the phonon propagation in silicon must be studied. A dedicated fast readout electronics, using channelized Digital Down Conversion for monitoring up to 12 MKIDs over a 100\,MHz bandwidth was developed.  Thanks to the fast ADC sampling and steep digital filtering, In-phase and Quadrature samples, having a high dynamic range, are provided at $\sim$2\,Msps. This paper describes the technical solution chosen and the results obtained.}


\begin{document}

\section{Introduction}
The microwave kinetic inductance detectors (MKID) are superconducting detectors, typically operated below 300\,mK. They consist of high-quality superconducting resonant circuits electromagnetically coupled to a transmission line and they are designed to resonate in the microwave domain \cite{Day,Mazin2,DoyleThesis,Baselmans}.
Their resonances lie between 1 to 10\,GHz and have loaded quality factors around $10^5$, corresponding to a typical bandwidth of about 10-100\,kHz.
As the MKIDs resonant frequencies can easily be controlled during manufacturing, by adjusting the capacitor associated to each inductance (hence creating a LC resonator), it is possible to couple a large number of MKIDs to a single transmission line without interference and therefore to perform a frequency-multiplexed readout \cite{Swenson}.

These detectors can perform either photon mediated detection (mm-wavelength astronomy\cite{Monfardini,Monfardini2011,Schlaerth}, ...) or phonon mediated detection (X-rays imaging \cite{Cruciani}, particle detector for cosmic-ray interactions, dark matter detection, ...). In the latter case, they can be used as a tool for characterizing phonon propagation in various materials.
The kinetic inductance ($\rm L_k$) is defined as the the inductance associated to the kinetic energy of the superconducting charge carriers (Cooper pairs). $\rm L_k$, and thus the resonant frequency, is sensitive to the Cooper pairs broken by the energy injected in the substrate by the incoming photon/particle. For the thin films used for MKID it can be demonstrated \cite{SwensonAPL2010} that the shift in frequency is proportional to the deposited energy. The pulse amplitude is thus a good parameter to use for building energy spectra.
Moreover, due to the metal layer thinness of the MKID ($<$20\,nm) which minimizes energy deposit, and to the low fill factor $<$1\% which minimizes the cross-section, the number of direct cosmic-ray interactions and the extra signal induced are very low.
This makes this kind of detector good candidates for next generation of space-borne mission. Indeed, during the Planck mission, the bolometers were severely affected by cosmic ray direct interactions \cite{2013arXiv1303.5071P},
therefore this issue has to be addressed for future missions. Within this context, the SPACEKIDS collaboration aims at optimizing these detector to further decrease their sensitivity to direct cosmic-ray interactions with the substrate.

\begin{figure}
\begin{center}
\includegraphics[angle=-90,width=0.8\textwidth]{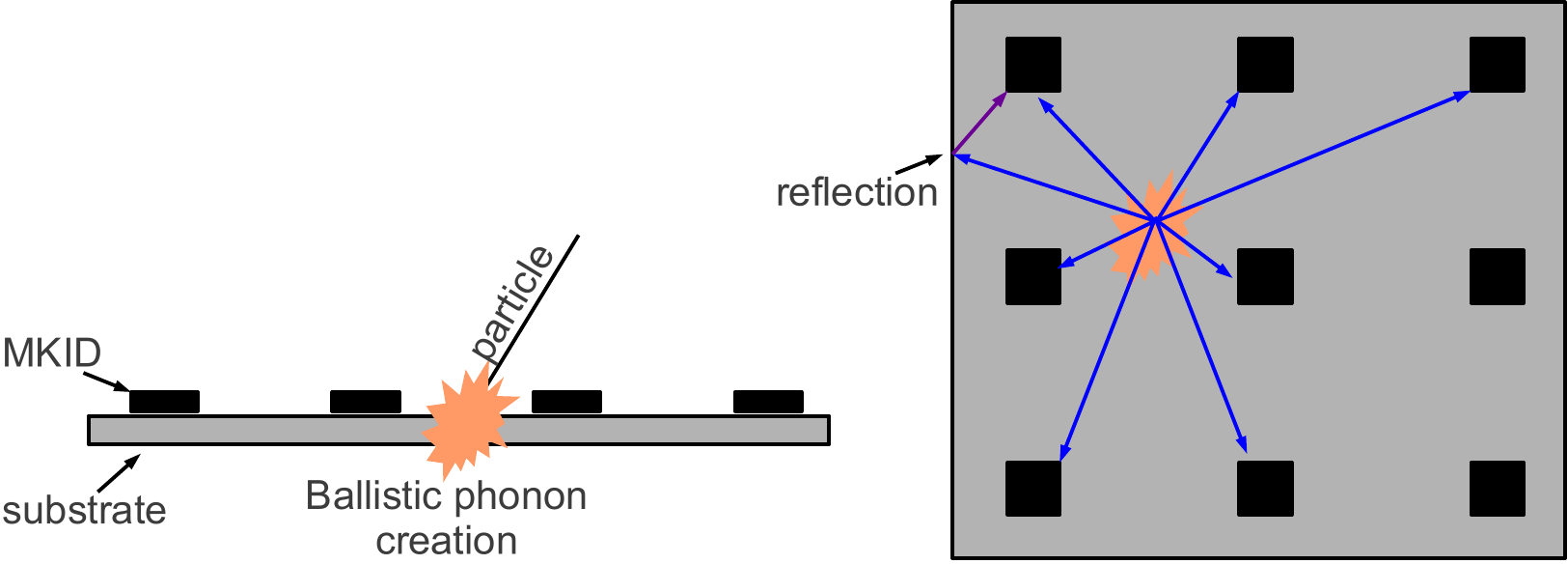}
\caption{A high energy particle interaction in the substrate is detected and localized by the time-resolved detection of the ballistic phonons produced. Direct and reflected phonons can be detected.}
\label{phononImaging}
\end{center}
\end{figure}

Consequently and for different goals that can be either time-resolved phonon-mediated detection of high-energy interactions from cosmic rays or optimization of the MKID array design, dedicated fast readout electronics are required to fully study phonon diffusion in the substrate.
See for example figure~\ref{phononImaging} where we show the interaction of a high energy particle in the MKID array substrate and how ballistic phonons
propagate.

\section{MKID instrumentation}
\begin{figure}
\begin{center} 
\includegraphics[angle=-90,width=0.7\textwidth]{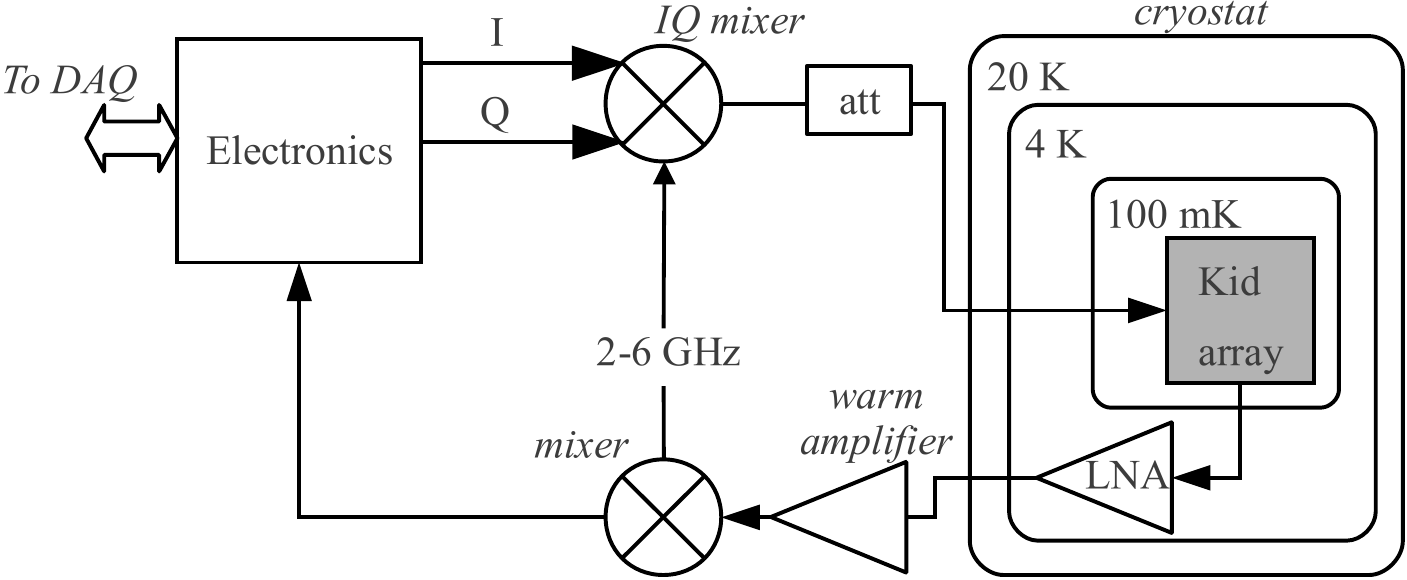}
\caption{Overview of the setup required to monitor a MKID array featuring the electronics generating the two frequency combs (each tone phase shifted by 90\textdegree\  between I and Q), the IQ up-converter, the programmable attenuator for power adjustment, the cryostat, the low noise boost amplifier, the down converter and the warm amplifier.}
\label{hardwareSetup}
\end{center}
\end{figure}

From an electrical point of view, an MKID array can be seen as a single transmission line equipped with a certain number of high quality factor resonators, each having a different self resonant frequency.
Figure~\ref{hardwareSetup} shows the typical instrumental setup.
The frequency comb, having each of its frequencies adjustable at a kHz resolution and provided in its In-phase (I) and Quadrature (Q) version, must be generated at baseband and up-converted to the frequency band of interest by hybrid mixers.
The excitation signal is then passed through the array at appropriate power level and eventually modified. 
Finally, the line output signal is amplified and down-converted back to baseband in order to be processed via channelized Digital Down Conversion (DDC) by the electronics.
Each resulting tone amplitude and phase can be determined, full details are available in \cite{Bourrion2011}.

\section{Electronics requirements}
In order to reach our goal, which is to be able to detect high energy particle interactions in the substrate and hence to have an investigation tool for MKID array design optimization, we need to match the following requirements:

\begin{itemize}
 \item \textbf{Achieve an acquisition rate of about 2\,Msps}: to cope with phonon propagation speed (typically in the order of mm/\textmu s) but also to cope with the typical MKID signal rise time (at 2\,GHz with Q=$10^5$ is a few \textmu s);
  
 \item \textbf{Be able to read-out an array of at least $3 \times 3$ MKIDs}: to be able to perform time-resolved 2D detection. Given the fact that MKID resonance frequencies are separated by a few MHz, this implies that the analog bandwidth must be larger than about 50\,MHz;
  
 \item \textbf{Use a triggered acquisition}: Since the array is monitored at high speed, only the useful events must be acquired. This requires an on-line trigger able to select the events of interest by triggering on the amplitude of the transmitted tone. The trigger level must be adjustable for individual resonators as each MKID does not have the same transfer function (in terms of transmission gain and dephasing).
 
 \item \textbf{Acquire a data frame of up to 1000 samples}, with the possibility to get up to 500 samples prior to trigger:  this is necessary for time-resolved detection and for phonon reflection discrimination. The possibility to record the baseline prior to the event detection allows baseline suppression and real time noise studies;
 
 \item \textbf{Have a measurement SNR better than $\sim$60\,dB}: the low-noise cold amplifier noise is followed by a typical amplifier chain gain of 60-65\,dB. Thus the best SNR achievable on this part is already limited to about 60\,dB.
\end{itemize}
\section{Hardware description}
\begin{figure}
\begin{center} 
\includegraphics[angle=-90,width=1\textwidth]{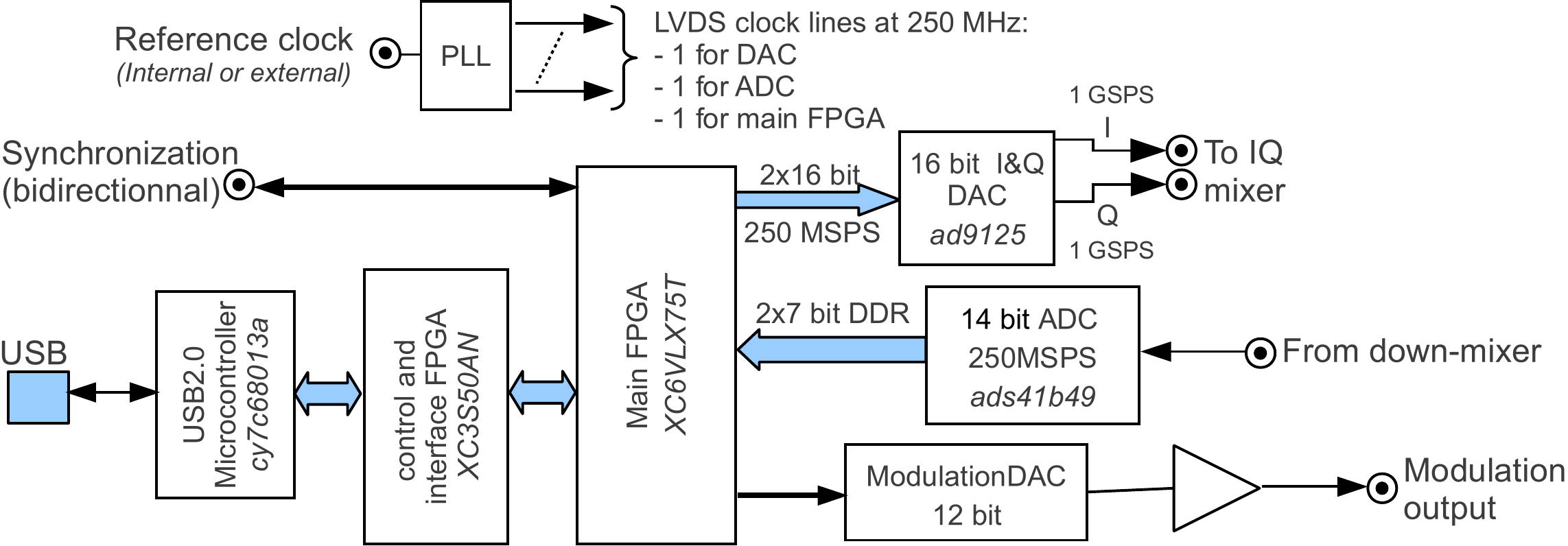}
\caption{Block diagram of the electronic board.}
\label{hardwareOverview}
\end{center}
\end{figure}

An overview of the board can be seen in figure~\ref{hardwareOverview}.
It is is built around a main FPGA (XC6VLX75T-2FF484C) that is in responsible  of generating the frequency comb and of performing signal processing.
The data acquisition and slow control interface to this FPGA is done by a USB2 micro-controller (CY7C68613) through another FPGA (XC3S50AN-2TQ144).
This latter combination allows the FPGA reconfiguration from the USB2 interface, which thus provides great flexibility. 

The digital version of the frequency comb generated by the main FPGA in its I and Q version is fed to a dual 16 bit DAC (AD9125) operated at 250\,Msps.
This DAC, featuring digital modulators and interpolators followed by very steep half-band filters, is able to provide analog samples at a rate of 1\,GSPS over a bandwidth of 100\,MHz.
Consequently, the mandatory output anti-aliasing analog filter constraints are greatly relaxed (aliasing occuring at 500\,MHz instead of 125\,MHz).

A fast ADC (TI ADS41B49), having a resolution of 14 bit and operated at 250\,Msps is used to sample the down-converted MKID array signal output and thus to provide the digital samples to the main FPGA for processing. 

To fulfill other applications, such as for instance mm-wave astronomy, this board was designed with some extra features. These are: the possibility to achieve multi-board synchronization, thanks to the use of the optional external clock reference and of the start signal input; the capability to control a radio-frequency source modulation with the slow DAC provided (up to a few kHz).
The required flavor of the board is simply selected by the firmware loaded in the main FPGA.

\section{Firmware description}
\begin{figure}
\begin{center} 
  \includegraphics[angle=0,width=0.8\textwidth]{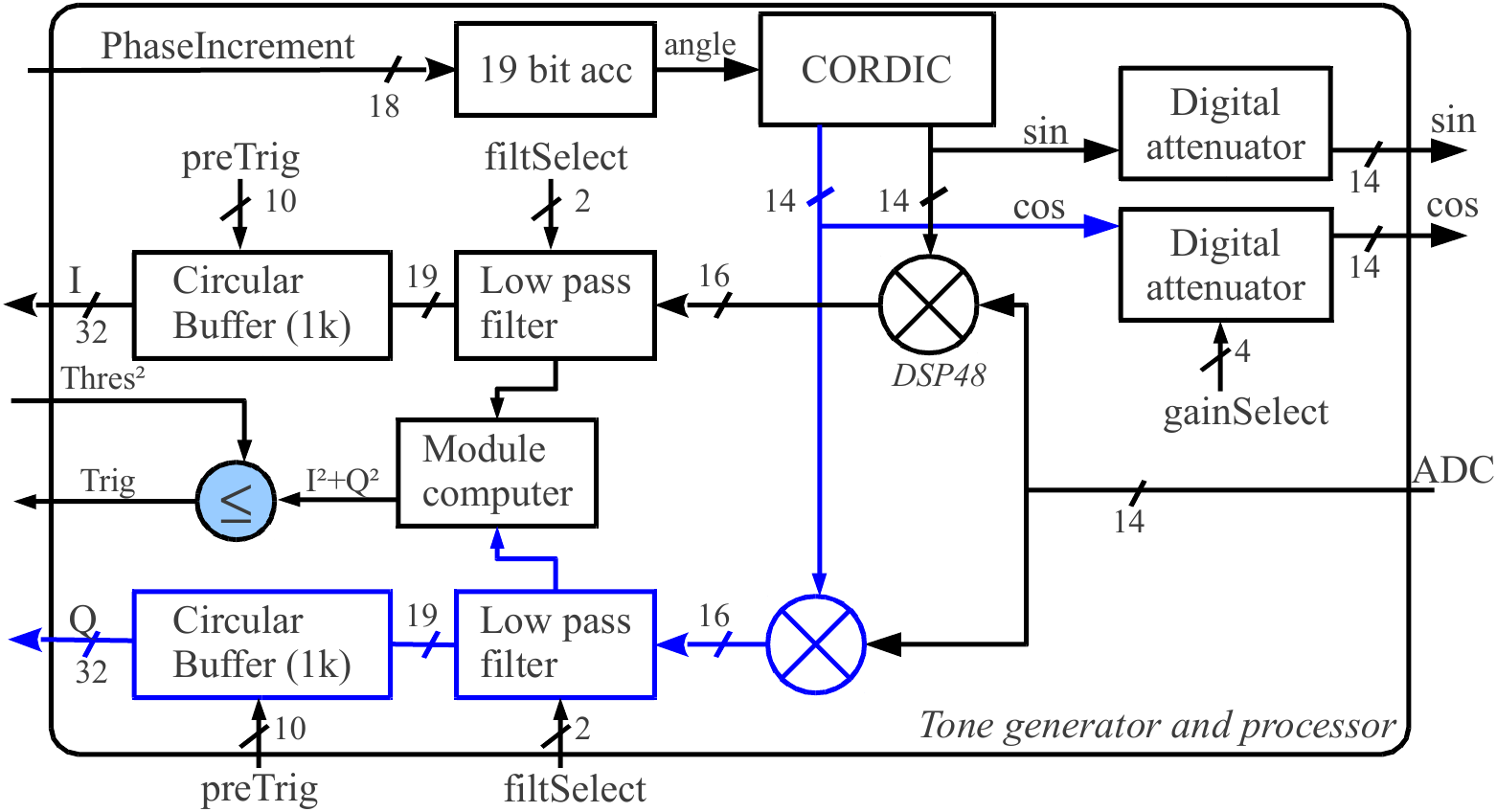}
\caption{Overview of the firmware building block: the tone manager; One of these is used per MKID instrumented.}
\label{toneManager}
\end{center}
\end{figure}

\begin{figure}
\begin{center} 
\includegraphics[angle=0,width=0.46\textwidth]{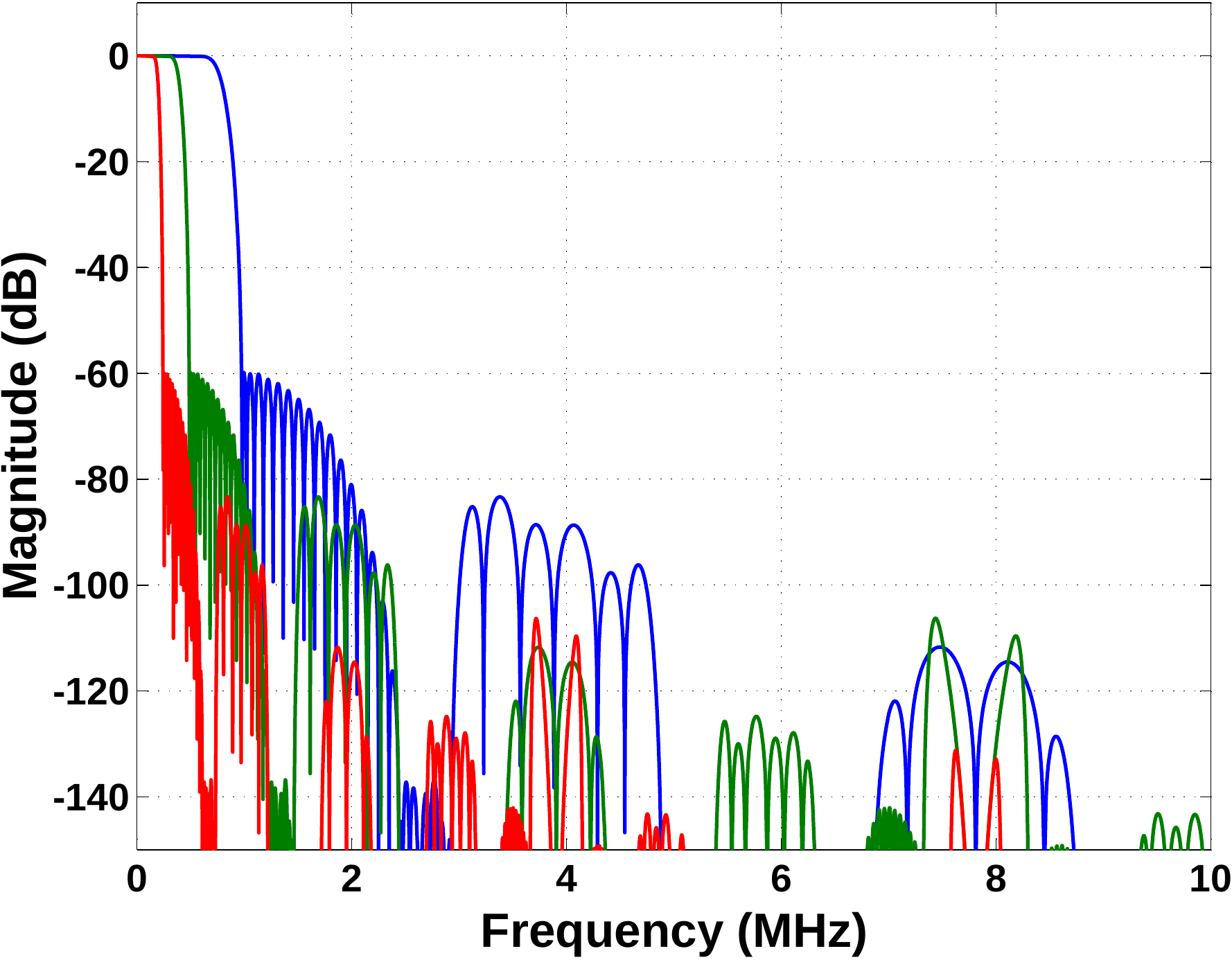}
\includegraphics[angle=0,width=0.46\textwidth]{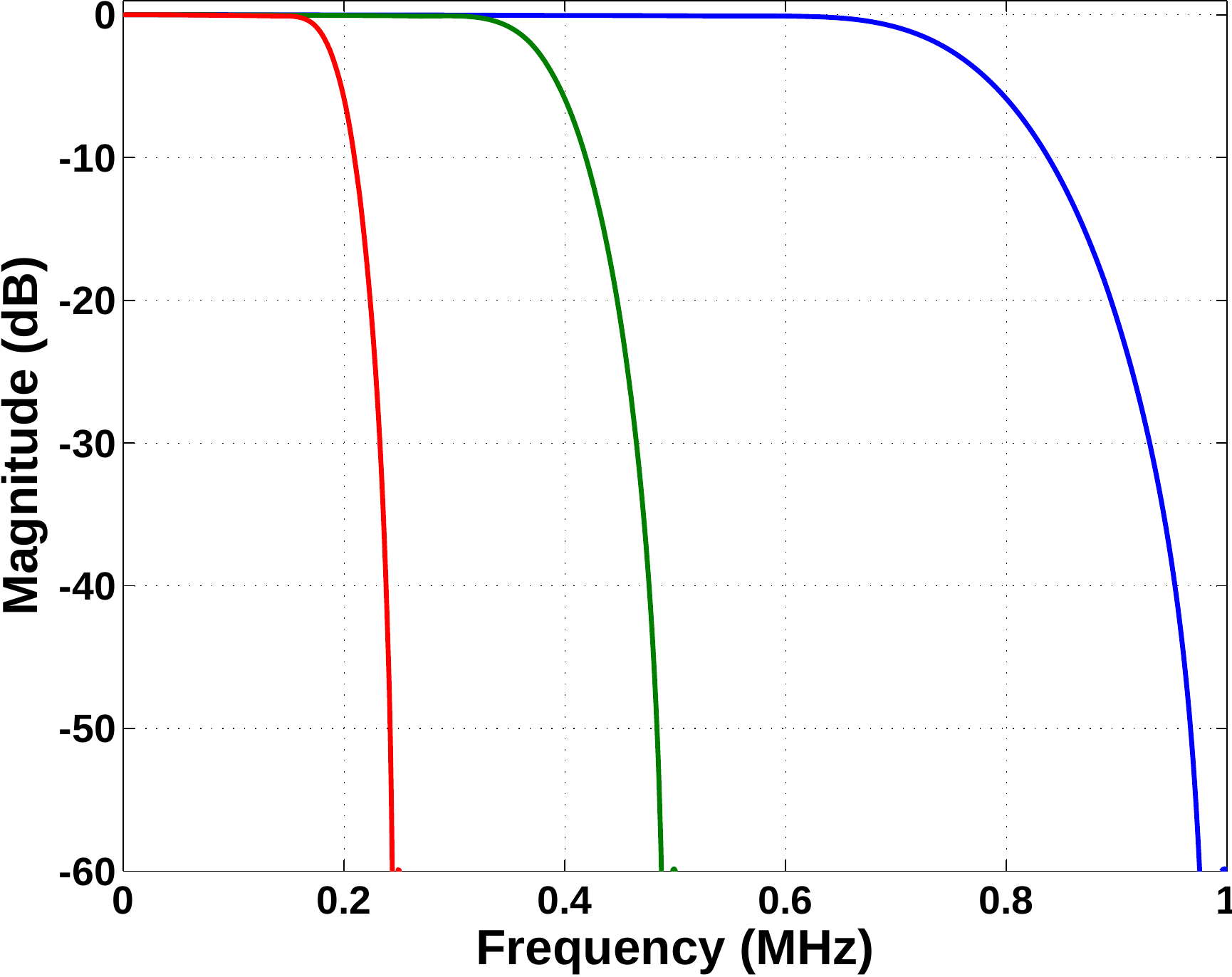}
\caption{Transfer function of the low-pass filters (red for 488\,kSPS, green for 0.97\,Msps, blue for 1.953\,Msps) used in the DDC. 10\,MHz frequency range (left) and zoomed on the low frequency range (right).}
\label{filterPlot}
\end{center}

\end{figure}

The building block of the firmware -- the tone manager -- is shown in figure~\ref{toneManager}.
It features a 14 bit resolution COordinate Rotation DIgital Computer (CORDIC)\cite{Volder} which is used to provide the sine and cosine waves for the DDC and for building the array excitation signal (possibility to apply attenuation in one eight steps). The CORDIC is a simple and efficient algorithm used to calculate sine and cosines without the use of hardware multipliers, its architecture makes it easy to implement in FPGA in a pipelined fashion, and thus allows high performance to be reached.
This CORDIC, driven by a 19 bit phase accumulator, allows a frequency resolution of $\rm 250\,MHz/2^{18}=953\,Hz$.
The multipliers (DSP48E) used in the DDC are followed by highly selective low pass filters (see figure~\ref{filterPlot}) and by circular buffers to provide data delays for pre-triggering adjustment.
The low-pass filter selectivity avoids crosstalk between MKID, provided their self-resonance frequencies are sufficiently separated and improve Signal to Noise Ratio (SNR) by reducing the white noise contribution in the stop-band. Another strong argument for the filter selectiveness is the ability to instrument without deformation the fast frequency shift in the pass-band.

\begin{figure}
\begin{center} 
\includegraphics[angle=90,width=0.85\textwidth]{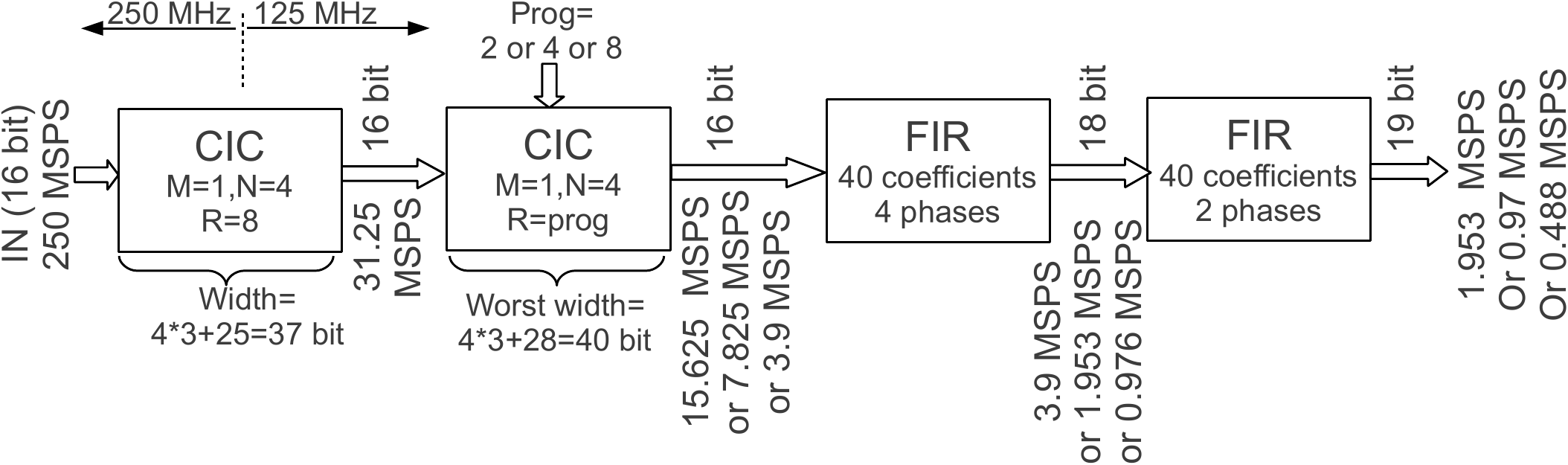}
\caption{Diagram of the low pass filter used in the DDC. Input sampling frequency is at 250\,Msps and after adequate filtering the output sampling rate ranges from  488\,ksps to 1.953\,Msps.}
\label{filterDesign}
\end{center}
\end{figure}

The DDC low pass filters are composed of four sub-filters, i.e two Cascaded Integrator-Comb (CIC) followed by two Finite Impulse Response (FIR) filters, see figure~\ref{filterDesign}.
The second stage of these filters has a selectable decimation rate, hence allowing a MKID sampling rate selectable between 1.953\,Msps and 488\,ksps.
To allow on-line triggering, the filter outputs are also used to compute in-line and at the selected sampling rate the square of the module ($I^2+Q^2$) that is constantly compared to the square of the amplitude thresholds (one per MKID).

For twelve MKIDs (24 filters), the filter resources usage versus their availability in the FPGA chosen are 36k/93k flip-flops, 21k/46k LUTs, 144/288 DSP48E.


\begin{figure}
\begin{center} 
\includegraphics[angle=-90,width=\textwidth]{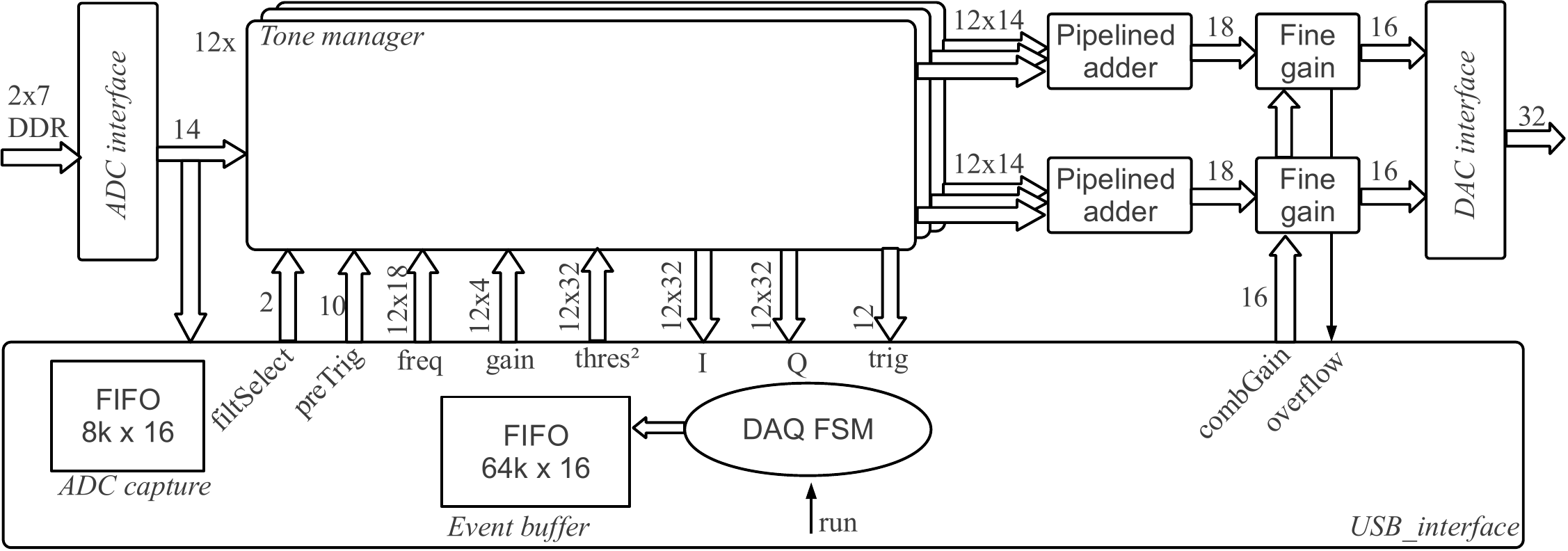}
\caption{Global overview of the firmware in the case of 12 MKIDs (e.g.  twelve instances of the --tone manager--).}
\label{firmwareOverview}
\end{center}
\end{figure}

A global overview of the firmware is shown in figure~\ref{firmwareOverview}.
It shows the ADC and DAC interfaces used to manage the external converters, and in between twelve instances of the --tone manager-- whose sine and cosine outputs are fed to pipelined adders that perform digital summations of the signals and hence build an I and Q digital version of the frequency comb.
These adders are followed by digital multipliers -- fine gain -- used to precisely adjust the resulting digital comb to the DAC dynamic range and thus to avoid an eventual digital overflow resulting in DAC clipping.
The gain can be adjusted in steps of 1/2\textsuperscript{16}, and it is adjusted so that full gain yields full scale for a single tone enabled, that is a gain of four.

A USB interface block is used to enable the data acquisition and the monitoring of the system through the USB2 microcontroller.
It contains a snapshot FIFO (8k words) used to directly capture the ADC interface output, a 64k words event recording FIFO which can fit a maximum of 1365 samples ($64k/((12 \times 2) \times 2)$), a finite state machine (FSM) which manages the data recording system, and an overflow monitoring (ADC and DAC) system.

For the FPGA chosen and for twelve MKIDs the resources usage versus availability in the FPGA chosen are 51k/93k FF, 38k/46k LUTs, 71/156 RAMB36, 156/288 DSP48.

\section{Performance and results}
To assess the performance of the electronics, two different kinds of measurements were performed.
In both cases, one of the DAC outputs was directly connected to the ADC input and a single frequency was generated at 51\,MHz at full scale by the DAC having an amplitude -12\,dBFS with respect to the ADC input range (due to DAC/ADC analog range mismatch).  

The first measurement used the snapshot FIFO to record 8192 data samples and to build the corresponding FFT spectrum to check that the board implementation did not degrade the ADC performance (noise floor at -100\,dBFS).

\begin{figure}
\begin{center} 
\includegraphics[width=0.75\textwidth]{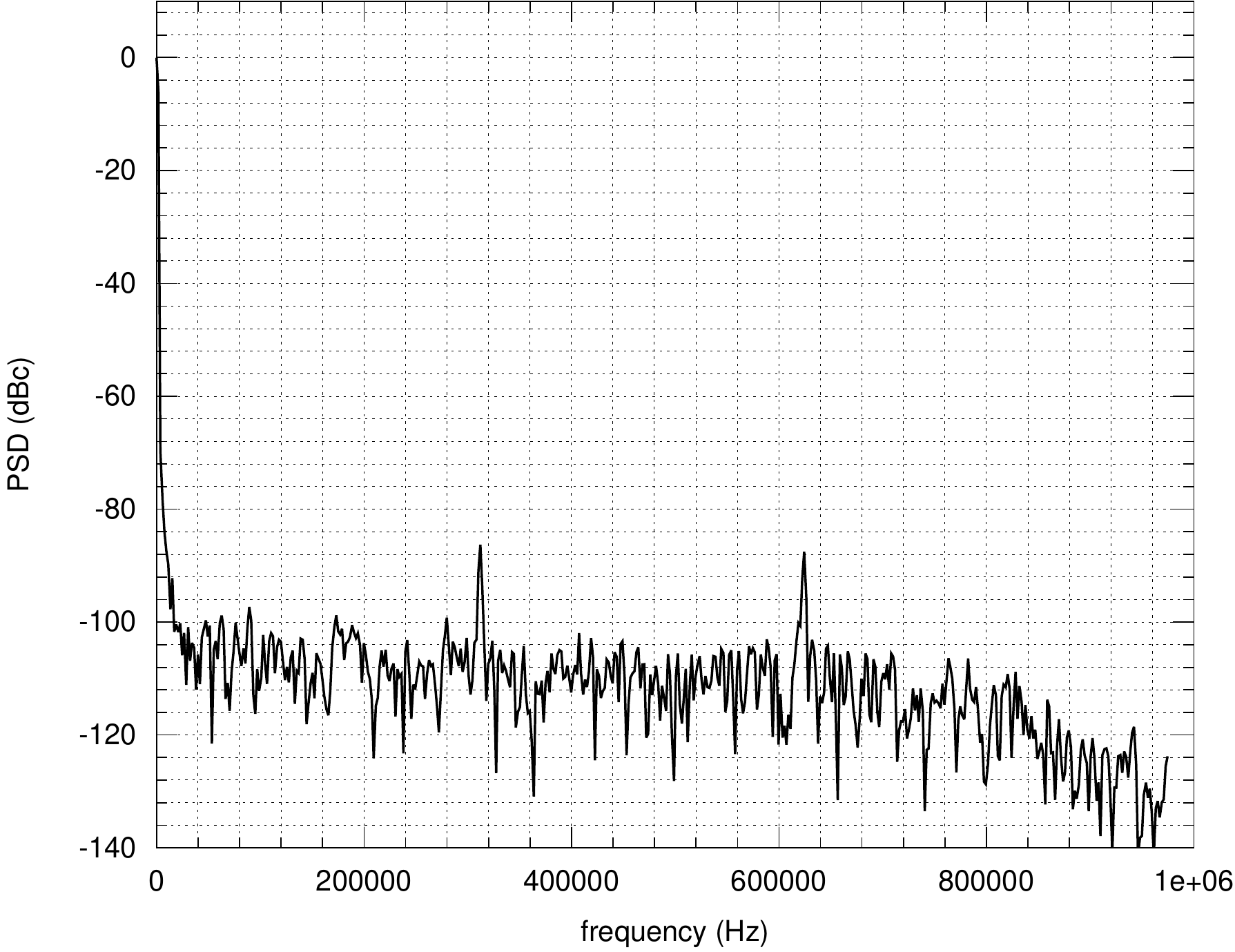}
\caption{Spectrum computed with 1024 DDC data recorded at 2\,Msps in a loop-back measurement with a single tone generated at 51\,MHz. Hanning filtering applied before FFT computation.}
\label{DDCnoisePlot}
\end{center}
\end{figure}

The second measurement used 1024 samples recorded via the event FIFO in the 1.953\,Msps case.
On these data, a Hanning filtering was applied and a typical power spectrum obtained via FFT is presented in figure~\ref{DDCnoisePlot}.
The high noise rejection (converted signal sits at 0\,Hz) can be observed: the noise floor is lower than -100\,dBc.
The two spikes sitting at multiples of 320\,kHz are explained by the fact that the DC/DC converter used to supply the FPGA core voltage is located too close to the pulse transformers used at the excitation DAC outputs.
An adequate shielding will be added to address this problem. It may be noticed that the last part of the spectrum (over 800\,kHz) shows the effect of the low pass filter turn on.

\begin{figure}
\begin{center} 
\includegraphics[width=0.27\textwidth]{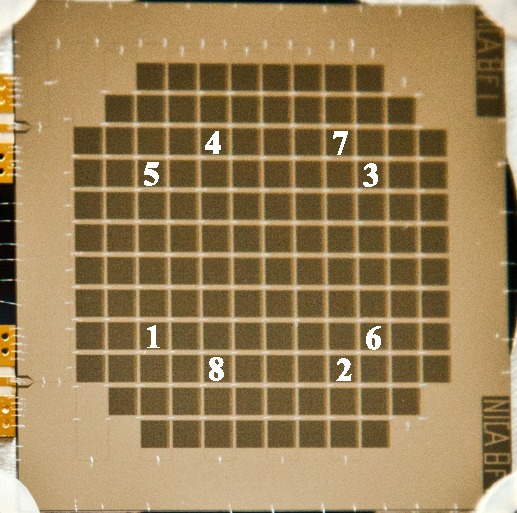}
\caption{Picture of the array used to perform functional tests. Only eight MKIDs were monitored as marked in the figure.}
\label{testArray}
\end{center}
\end{figure}

\begin{figure}
\begin{center} 
\includegraphics[width=0.49\textwidth]{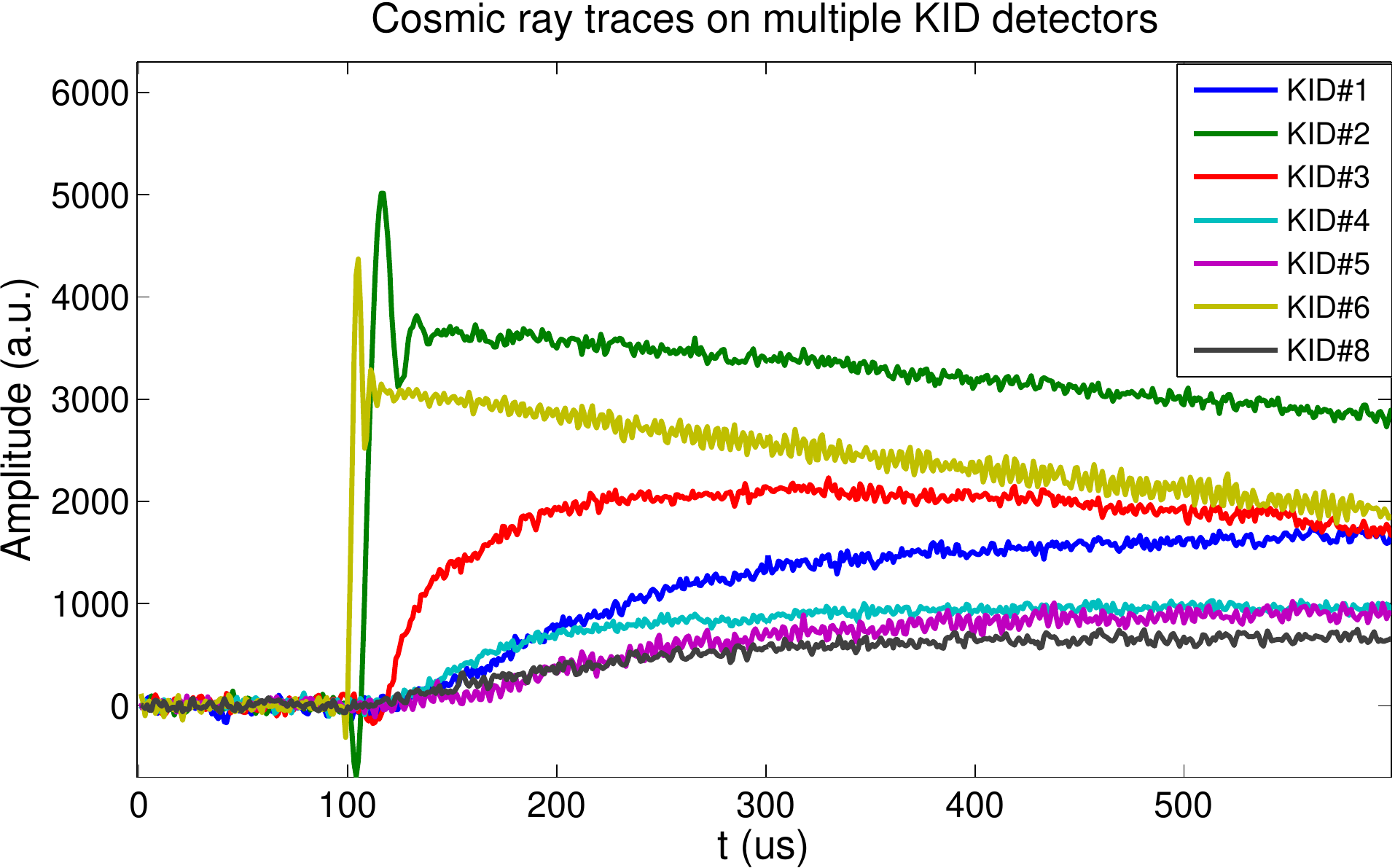}
\includegraphics[width=0.49\textwidth]{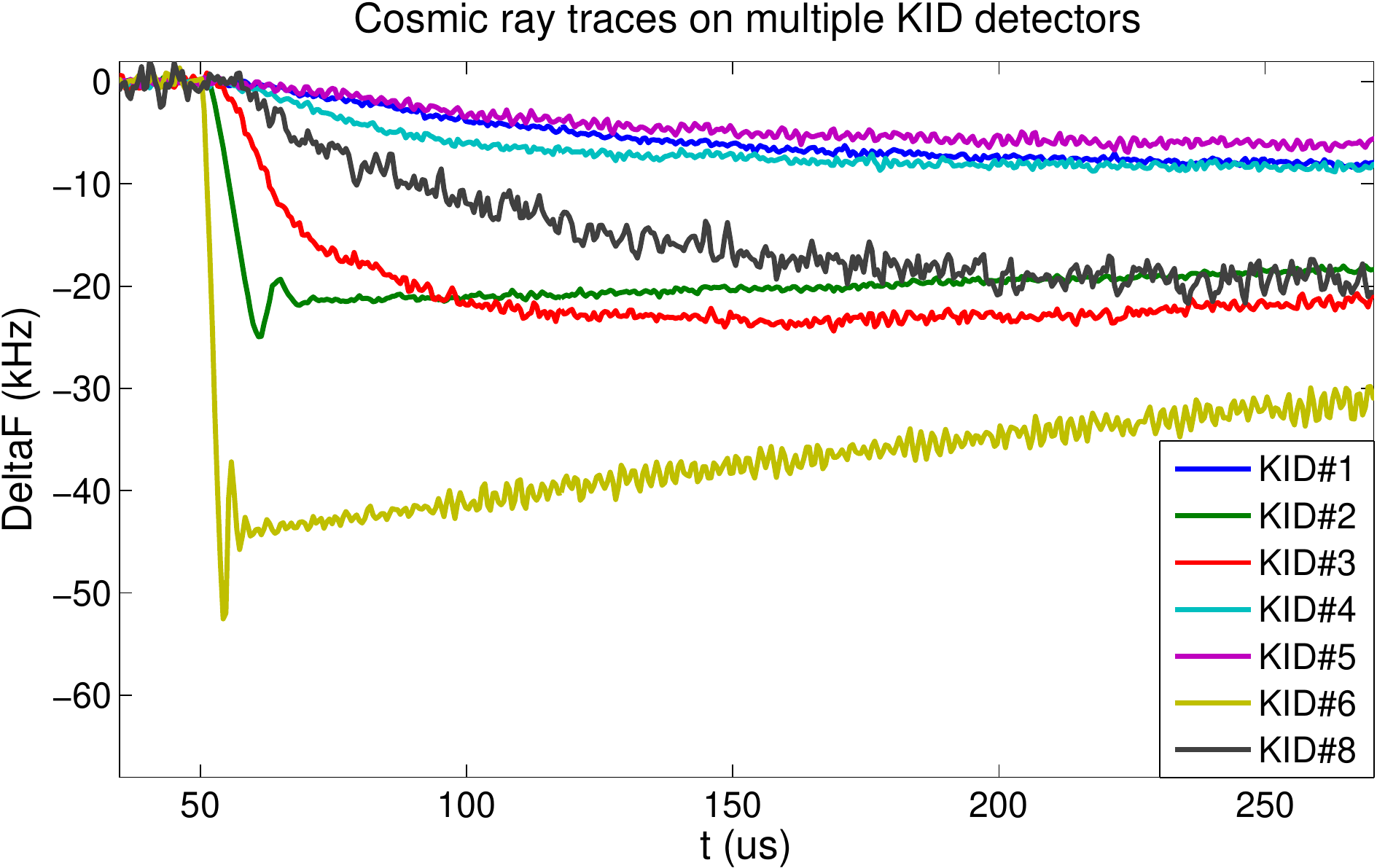}
\caption{Signal amplitude (left) and frequency shift (right), as measured by each active MKID, for the hit of a cosmic ray on the array substrate.
For the amplitude the baseline is removed for clarity. Frequency variations show the real amount of deposited energy.}
\label{event}
\end{center}
\end{figure}

As an illustration of the typical use of the electronics board presented in this paper
we performed a test on eight pixels of an MKID array (see figure~\ref{testArray})
The main goal of this test was to detect cosmic ray hitting the array substrate.
The sampling was performed at 1.953\,Msps and 600 data samples were recorded.
Figure~\ref{event} shows Signal amplitude (left) and frequency shift (right), $\Delta f$, as measured by each active MKID, for the hit of a cosmic ray on the array substrate. 
We observe for  MKID 2 and 6 a fast rise and then a long time-constant decay of the amplitude signal, and equivalently a fast decrease and then long time constant increase of the frequency variation.
For the other MKIDs the increase in amplitude (respectively decrease in $\Delta f$)
is much slower.
Furthermore, from the frequency variation data, that are proportional to the number of generated quasi-particles in the superconductor, we can see that the sampling rate is large enough to see the time difference between the detection and thus that the cosmic-ray impact occurred close to pixel 2 and 6. 
The physical interpretation will be reported in future publications.
Given the fact that each MKID resonance depth is different, the amplitude plots are shown with baseline removed for clarity.

\section{Summary and conclusions}
In this paper we have presented the design and implementation of a fast electronic board, with acquisition rate up to 1.953\,Msps, dedicated to the readout of arrays of MKIDs (up to 12 pixels).
We have demonstrated that a triggered acquisition of 1365 data samples is possible with SNR better than about 60\,dB.
In real conditions, with an eight pixels MKID array, we have been able to distinctively detect, in each pixel, the signal from a cosmic-ray hitting the substrate of the array.
This confirms that the electronic board detailed here is a powerful tool for investigating cosmic-rays interactions in MKID arrays that is a key issue to be solved by the SPACEKIDS collaboration.
Such a tool will be an asset for the characterization and optimization of MKID arrays for future space missions in the millimeter wave domain.
Finally, potential applications of this electronics can be for instance 10-100\,keV X-ray large field (10\,cm) imaging and rare event searches (e.g. double-beta decay), dark matter, etc.

\section*{Acknowledgements}
This work has been supported as part of a collaborative project, SPACEKIDS, funded via grant 313320 provided by the European Commission under Theme SPA.2012.2.2-01 of Framework Programme 7. 


\begin{thebibliography}{9}
\bibitem{Day} P. K. Day, Nature 425, Issue 6960, pp 817-821 (2003)
\bibitem{Mazin2} B. Mazin et al, $9^{th}$ International Workshop on Low Temperature Detectors. AIP Conference Proceedings, Volume 605, pp. 309-312 (2002)
\bibitem{DoyleThesis} S. Doyle, Lumped Element Kinetic Inductance Detectors. PhD thesis, Cardiff University, United Kingdom, 2008
\bibitem{Baselmans} J.~Baselmans, Kinetic inductance detectors, JLTP, online first (2012), DOI:10.1007/s10909-011-0448-8
\bibitem{Swenson} Swenson L.~J., et al., 2009, AIPC, 1185, 84 

\bibitem{Monfardini} A. Monfardini et al 2010, NIKA: A millimeter-wave kinetic inductance camera, Astronomy and Astrophysics, Volume 521, id.A29 (2010), \href{http://arxiv.org/abs/1004.2209}{arXiv:1004.2209}
\bibitem{Monfardini2011} A. Monfardini et al 2011, The new NIKA: A dual-band millimeter-wave kinetic inductance camera for the IRAM 30-meter telescope, ApJS 194 24 (2011)	\href{http://arxiv.org/abs/1102.0870}{arXiv:1102.0870}
\bibitem{Schlaerth} A.J.~Schlaerth et al, the status of MUSIC: A multicolor sub/millimeter MKID instrument, JLTP, online first (2012), DOI:10.1007/s10909-012-0541-7
\bibitem{Cruciani} A.~Cruciani, X-Ray Imaging Using LEKIDs, JLTP Vol. 167, Issue 3-4, pp 311-317 (2012), DOI:10.1007/s10909-012-0549-z
\bibitem{SwensonAPL2010} Swenson L.~J., et al., High-speed phonon imaging using frequency-multiplexed kinetic inductance detectors, Appl. Phys. Lett. 96, 263511 (2010)

\bibitem{2013arXiv1303.5071P} {Planck Collaboration}:  P.~A.~R. Ade et al, Planck 2013 results X. Energetic particle effects: characterization, removal, and simulation, A\&A submitted, 2013,  \href{http://arxiv.org/abs/1303.5071}{arXiv:1303.5071}
\bibitem{Bourrion2011} O. Bourrion et al 2011, Electronics and data acquisition demonstrator for a kinetic inductance camera, 2011 JINST 6 P06012, \href{http://arxiv.org/abs/1102.1314}{arXiv:1102.1314}
\bibitem{Volder} J. Volder, The CORDIC Trigonometric Computing Technique, IRE Transactions on Electronic Computers, pp330-334, September 1959

\end{thebibliography}
\end{document}